\documentclass[12pt]{iopart} 

\pdfoutput=1

\usepackage{iopams}  
\usepackage[english]{babel}      

\newcommand{\text}[1]{\ensuremath{{\rm #1}}} 

\usepackage{graphicx}

\usepackage{subfigure}

\newcommand{\figref}[1]{Fig.~\ref{#1}}

\usepackage{color}
\usepackage[normalem]{ulem} 
\DeclareMathAlphabet{\Ibb}{U}{msb}{m}{n}
\newcommand   {\IC}{{\ensuremath{\Ibb C}}}
\newcommand   {\II}{{\ensuremath{\Ibb I}}}

\newcommand{\matone}{\ensuremath{\boldsymbol{\mathsf{1}}}}

\renewcommand{\tr}{\ensuremath{{\textrm{tr}}}}
\newcommand{\dev}{\ensuremath{\textrm{dev}}}

\newcommand{\Ba}{{\boldsymbol{\mathnormal a}}}
\newcommand{\Be}{{\boldsymbol{\mathnormal e}}}
\newcommand{\Bl}{{\boldsymbol{\mathnormal l}}}
\newcommand{\Bu}{{\boldsymbol{\mathnormal u}}}
\newcommand{\Br}{{\boldsymbol{\mathnormal r}}}

\newcommand{\superscr}[1]{\ensuremath{{}^{\rm #1}}}

\newcommand{\Bsigma }{\ensuremath{\boldsymbol\sigma}}

\newcommand{\Bve    }{\ensuremath{\boldsymbol\varepsilon}}
\newcommand{\ve     }{\ensuremath{           \varepsilon}}
\newcommand{\Bvepl  }{\ensuremath{\boldsymbol\varepsilon\superscr{pl}}}

\newcommand{\onetwo}{{\textstyle\mthin\frac{1}{2}\mthin}}          
\newcommand{\onethree}{{\textstyle\mthin\frac{1}{3}\mthin}}          
\newcommand{\twothree}{{\textstyle\mthin\frac{2}{3}\mthin}}          
\newcommand{\threetwo}{{\textstyle\mthin\frac{3}{2}\mthin}}          
\def\mthin{\mkern\thinmuskip}

\begin{document}

\title[Deformation Patterns and Surface Morphology in a Minimal Model of \ldots]%
{Deformation Patterns and Surface Morphology in a Minimal Model of Amorphous Plasticity}

 \author{Stefan Sandfeld${}^1$  and  Michael Zaiser${}^{1,2}$ }

 \address{ ${}^1$Institute of Materials Simulation (WW8),
           Friedrich-Alexander-University Erlangen-N\"urnberg,
           Dr.-Mack-Str. 77, 90478 F\"urth, Germany.%
           \\
 	  	   ${}^2$School of Engineering, 
 	  	   Institute for Materials and Processes, 
 	  	   The University of Edinburgh, 
 	  	   Edinburgh EH93JL, UK}
 	  	   
 \ead{stefan.sandfeld@fau.de}

\begin{abstract}
We investigate a minimal model of the plastic deformation of amorphous materials. The material elements are assumed to exhibit ideally plastic behavior (J2 plasticity). Structural disorder is considered in terms of random variations of the local yield stresses. Using a finite element implementation of this simple model, we simulate  the plane-stress deformation of long thin rods loaded in tension. The resulting strain patterns are statistically characterized in terms of their spatial correlation functions. Studies of the corresponding surface morphology reveal a non-trivial Hurst exponent $H \approx 0.8$, indicating the presence of long range correlations in the deformation patterns. The simulated deformation patterns and surface morphology exhibit persistent features which emerge already at the very onset of plastic deformation, while subsequent evolution is characterized by growth in amplitude without major morphology changes. The findings are compared to experimental observations. 
\end{abstract}

\maketitle


\section{Introduction}
\label{intro}
The plastic deformation of amorphous materials is characterized by spontaneous localization of plastic flow (formation of shear bands) \cite{Steif82}. The formation and interaction of shear bands may give rise to complex deformation patterns with long range (possibly fractal) correlations \cite{Poliakov94,Sun11}. Various models have been proposed to describe the interplay between local randomness and long-range stress re-distribution which forms the physical basis of these patterns. Roux and Hansen \cite{Roux92} consider an ideally plastic random fuse model, which is mathematically equivalent to deformation in pure anti-plane shear. By mapping the model on the optimal path/directed polymer problem, they demonstrate that in this case plastic activity localizes on a single slip line with self-affine morphology. Vandembroucq, Roux, and co-workers \cite{Talmali12,Talamali11} consider plane strain deformation with macroscopically uniaxial loading. They constrain the plastic strain tensor to be aligned with the macroscopic stress, which again leads to a scalar problem. In their model, deformation is assumed to proceed in discrete steps which occur once the local stress exceeds a deformation threshold which fluctuates randomly with space and strain. This model is largely equivalent to a model of crystal plasticity proposed by Zaiser and Moretti \cite{Zaiser05,Zaiser07} who consider shear deformations on a single slip system to occur in response to the respective resolved shear stress, and account for random variations of the microstructure in terms of local flow stress variations. Besides investigation of temporal aspects of plastic flow (intermittency and avalanches \cite{Talamali11,Zaiser05,Zaiser07}, these models have been used to study the morphology of shear band patterns \cite{Talmali12} as well as the associated surface morphology \cite{Zaiser05}.

Experimental investigations of the surface morphology of deformed samples \cite{Zaiser04,Wouters05,Wouters06,Schwerdtfeger10} have focused on 
single- and polycrystalline materials, where both simulation and experiment indicate the presence of long range correlations in the strain patterns as revealed by a non-trivial Hurst exponent $H\approx 0.8$ of the surface of the deformed samples. Such correlations have also been directly studied by investigating the surface patterns of slip lines. Kleiser and Bocek \cite{Kleiser86} studied the point set obtained by determining slip line intercepts with a line perpendicular to the slip plane, for which they determined a fractal dimension of $D \approx 0.5$. This is consistent with a fractal dimention of the slip line pattern of $D \approx 1.5$ and, under the additional assumption that  that all slip lines carry approximately equal strains, with a Hurst exponent of the surface profiles of $H \approx 0.75$. For metallic glasses, an investigation of Sun and Wang \cite{Sun11} also points towards fractal behavior of the deformation patterns, with a fractal dimension of about $1.5$ for the surface pattern of shear bands. Under the same assumption that all shear bands are of approximately equal strength, this again leads to the expectation of a surface profile Hurst exponent of around 0.75 \cite{Zaiser04}. 

All the above mentioned models have in common that plastic deformation is characterized by a single, scalar strain variable. Thus, effects of local stress multi-axiality cannot be taken into account. However, in case of materials which deform in an intrinsically heterogeneous and localized manner, such effects cannot be avoided even if the material is uniaxially loaded on the macroscopic (specimen) scale. In crystals, physical constraints associated with the lattice structure require shear deformation to occur on discrete slip systems. In amorphous materials, on the other hand, there are a priori no preferred slip directions, and one would therefore expect the local strain axis to fluctuate in space and/or time. In the present paper, we follow the approach of Hansen and Roux \cite{Roux92} and Burnley \cite{Burnley2013} in assuming that the material behavior can be approximated as ideally plastic, and in describing disorder in terms of random local variations of the yield stress. To determine the local shearing directions, we assume a standard associated flow rule which naturally accounts for effects of stress multi-axiality and for local fluctuations in the direction of principal stresses. We investigate the relevance of multiaxial stresses and analyze the emerging deformation patterns as well as the associated surface morphology.

\section{The model}
\label{sec:model}

\subsection{System and geometry} 
\begin{figure}
  \centering
  \hbox{}\hfill
  \subfigure[system geometry]{\includegraphics[height=0.5\textwidth]{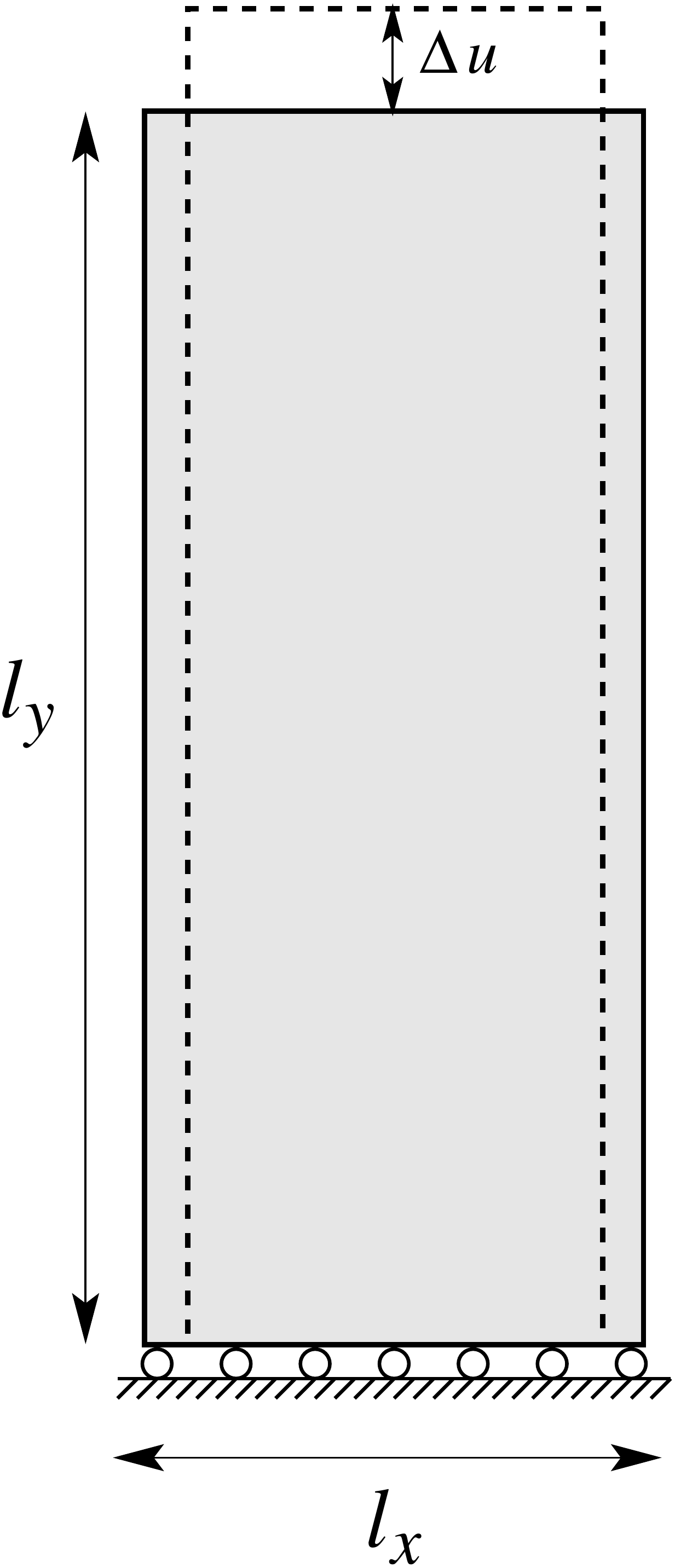}}
  \hfill
  \subfigure[yield stress distribution]{\includegraphics[height=0.5\textwidth, viewport=0 -30 265 420, clip]{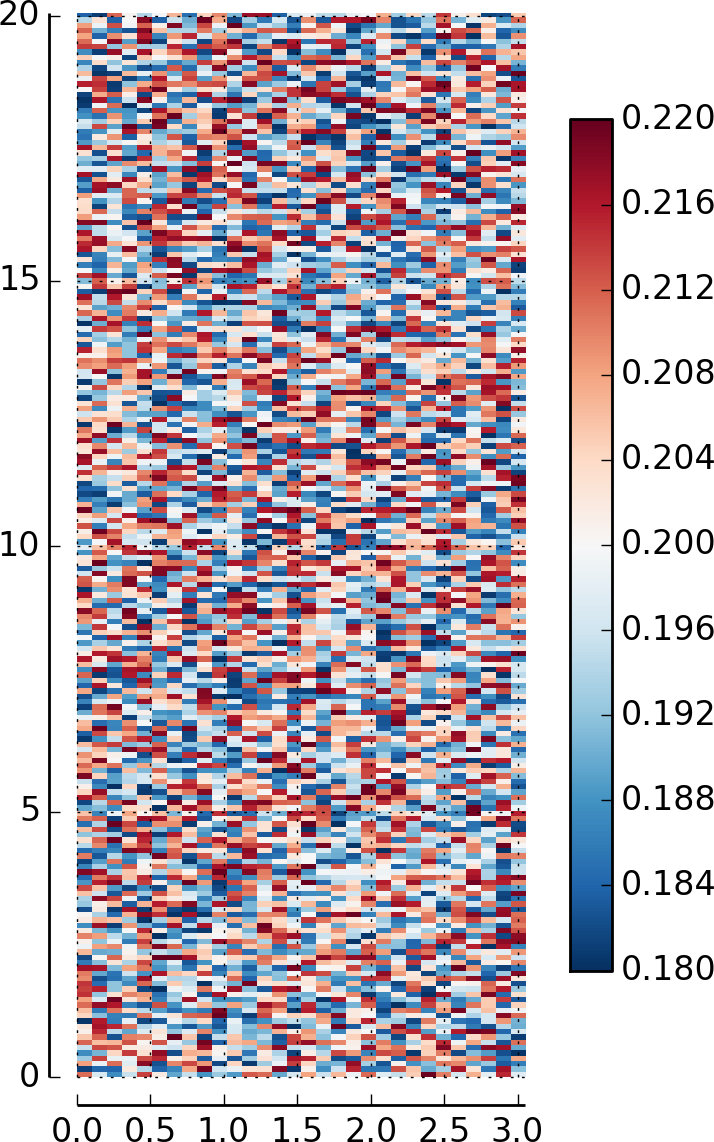}}
  \hfill\hbox{}
  \caption{\label{fig:rnd}%
      System geometry (left) and  example for an initial yield stress distribution (right). Each element has a constant but 
      randomly chosen yield stress (based on a uniform random distribution), which 
      fluctuates by $\pm\Delta\sigma=0.02$ around the mean yield stress value $\sigma_0=0.2$.
  }
\end{figure}

We consider long rectangular samples where initially the smaller side (which is parallel to the $x$ direction) has length $l_x=3$ and the longer side (which is parallel to the $y$ direction) has length $l_y=20$ (\figref{fig:rnd} (a)). All quantities are taken as dimensionless quantities. We assume that the deformation is plane stress in the $x-y$-plane, hence, the system can be considered of very small extension in $z$ direction. The case of plane strain, where the deformation into $z$ direction is assumed to be homogeneous, has also been investigated and we find no qualitative differences between both cases.
%
%
%
We simulate tensile deformation along the $y$ direction by prescribing a vertical displacement $u_y$ at the top face, while keeping the lower face vertically fixed. In addition, the lower left corner of the sample is constrained in horizontal direction. All other faces are unconstrained and may deform freely. The maximum normal strain due to the prescribed vertical displacement is $\max\varepsilon_{yy}=0.5$.

\subsection{Definitions}
For the description of our material model we use the following definitions: 
$\delta_{ij}$ denotes the Kronecker delta, which is $1$ for $i=j$ and $0$ otherwise; the second-order symmetrical identity tensor is defined as $\matone:=\delta_{ij}\Be_i\otimes\Be_j$, where the $\Be_k$ are basis vectors of the Cartesian coordinate system that we assume throughout;  $\otimes$ denotes the tensor product. $\II$ is the fourth-order symmetric identity tensor defined by $\II:=\onetwo(\delta_{ik}\delta_{jl}+\delta_{il}\delta_{jk})\Be_i\otimes\Be_j\otimes\Be_k\otimes\Be_l$. The trace of the rank-2 tensor $\Ba=a_{ij}\Be_i\otimes\Be_j$ is the sum of its diagonal elements, $\tr(\Ba)=\sum_i a_{ii}$, and the deviatoric part of $\Ba$ is defined as $\dev(\Ba)=\Ba-\onethree\tr(\Ba)\matone$. As a consequence, each rank-2 tensor can be decomposed into a deviatoric (or distortional) part $\dev(\bullet)$, and a volumetric part $\onethree\tr(\bullet)\matone$.

\subsection{$J_2$ plasticity}
Deformation of amorphous materials is driven by shear stresses, and deformation is found to be approximately volume conserving. Therefore, we choose a material model for which volumetric deformations do not contribute to the irreversible plastic deformation, and which therefore only depends on deviatoric quantities. Furthermore, any constitutive law should be frame invariant (i.e. objective), which means that the yield criterion can only depend on invariants of the (deviatoric) stress tensor because those invariants are by definition independent of the chosen coordinate system. The simplest criterion fulfilling this requirement considers only the $J_2$ invariant of the stress tensor which is defined as $J_2:= \onetwo\dev(\Bsigma)\!:\!\dev(\Bsigma)$ where '$:$' denotes the double contraction between two rank-2 tensors. This leads to the \emph{von Mises} or associated $J_2$ plasticity model (e.g. \cite{Simo1997}). In this model, the local material response is assumed to be linearly elastic -- ideally plastic with an elastic tangent modulus given by 
\begin{eqnarray}
	\IC^\text{e}=\kappa\matone\otimes\matone - 2\mu(\II-\onethree\matone\otimes\matone),
\end{eqnarray}
where $\kappa$ and $\mu$ are the bulk modulus and the shear modulus, respectively. 
According to the von Mises yield criterion, plastic deformation occurs instantaneously once the equivalent stress
\begin{eqnarray}
	\sigma_\text{eq}:=\sqrt{3J_2}=\sqrt{ \threetwo\text{dev}({\Bsigma}):\text{dev}({\Bsigma}) }
\end{eqnarray}
reaches a threshold, the yield stress  $\sigma_\text{y}$. The admissible stress states are defined by a yield criterion function $f(\sigma_\text{eq}):=  \sigma_\text{eq} - \sigma_\text{y}$ which is smaller than zero if the stress state is purely elastic, and which is zero on the {\em yield surface} where the yield stress is reached and plastic flow occurs. Stress states above the yield stress are not admissible. If in the plastic regime isotropic hardening occurs, then the yield stress is not a constant anymore and obeys the relation 
\begin{equation}  \label{eq:yield_function}
	\sigma^\text{p}_\text{y}(\varepsilon_\text{eq})=\sigma_\text{y0}+ \Theta\varepsilon^\text{p}_\text{eq},
\end{equation}
where $\Theta$ is a material parameter, the isotropic hardening modulus, and
\begin{equation}
\varepsilon^\text{p}_\text{eq}:=\sqrt{ \twothree\text{dev}(\Bvepl):\text{dev}(\Bvepl) } 
\end{equation} 
is the equivalent plastic strain which is work conjugate to the equivalent stress. The amount of plastic flow per load increment is governed by the \emph{Kuhn-Tucker conditions}
\begin{equation}
	f\leq 0,\quad \lambda\geq 0,\quad \lambda f=0,
\end{equation}
where $\lambda$ is the plastic multiplier, which governs the rate of plastic slip and which can be determined from the consistency condition $\lambda \dot{f}(\sigma_\text{eq})=0$ (if ${f}(\sigma_\text{eq})=0$). For details on the algorithmic details for iteratively solving this problem in a three-dimensional setup we refer the reader to e.g. \cite{Simo1997}.

\subsection{Material parameters, numerical discretization and solution} 
In our model, we assume that the bulk modulus is $\kappa=0.83$ and the shear modulus $\mu=0.38$, approximately giving a Young's modulus $E = 1$ and Poisson number $\nu = 0.3$. We  furthermore assume that the yield stress fluctuates spatially within the specimen: we divide the sample into (30 x 200) quadratic elements of equal size $\Delta x=\Delta y=0.1$ and choose the yield stress $\sigma_{y0}$, which is assumed constant within each element, randomly from a uniform distribution over the interval
\begin{equation}
	\sigma_\text{y0} \in [ \overline{\sigma}_\text{0} - \Delta\sigma,\overline{\sigma}_\text{0} + \Delta\sigma]
\end{equation}
with $\overline{\sigma}_0 = 0.2$ the mean value and $\Delta\sigma$ the maximum deviation of the local yield stresses. We use two yield stress distributions representing different degrees of disorder with  (A) $\Delta\sigma=0.02$ and (B) $\Delta\sigma=0.15$. For numerical reasons we add a small isotropic hardening contribution $\Theta \ll E$ which affects the deformation behavior in a negligible way.

\begin{figure}
	\centering

	\subfigure[$u_x$]{ 
	\includegraphics[height=0.45\textwidth]{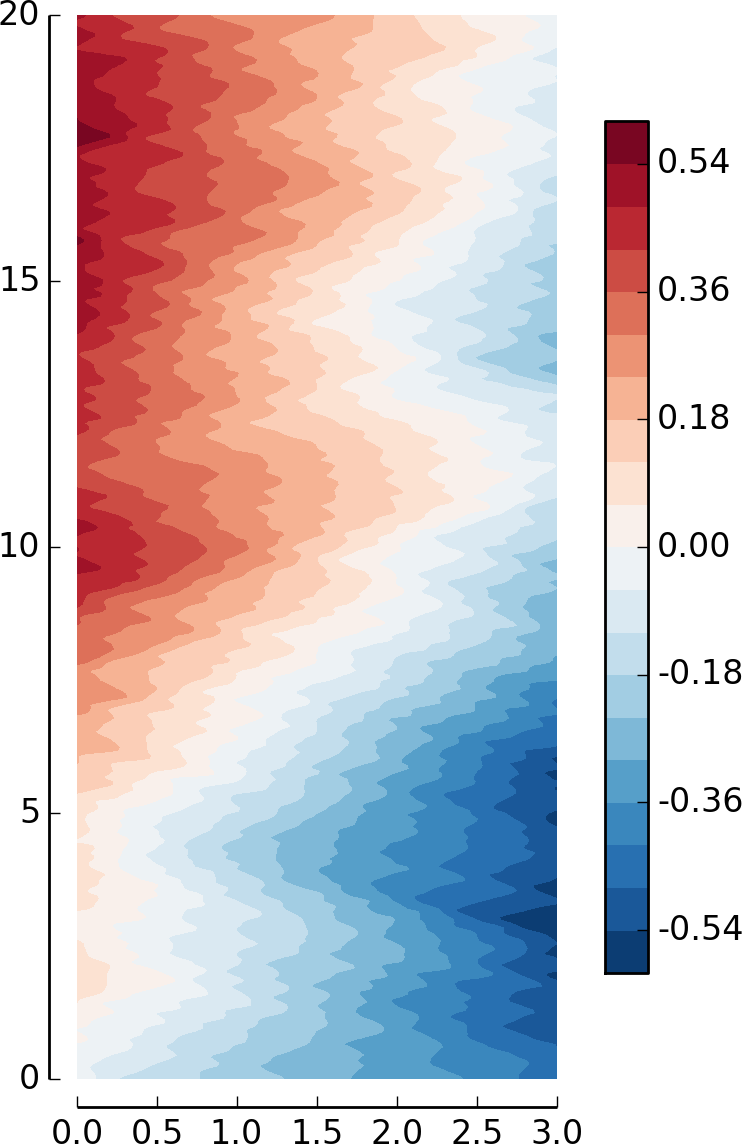}}\hfill%
	\subfigure[$u_y$]{ 
	\includegraphics[height=0.45\textwidth]{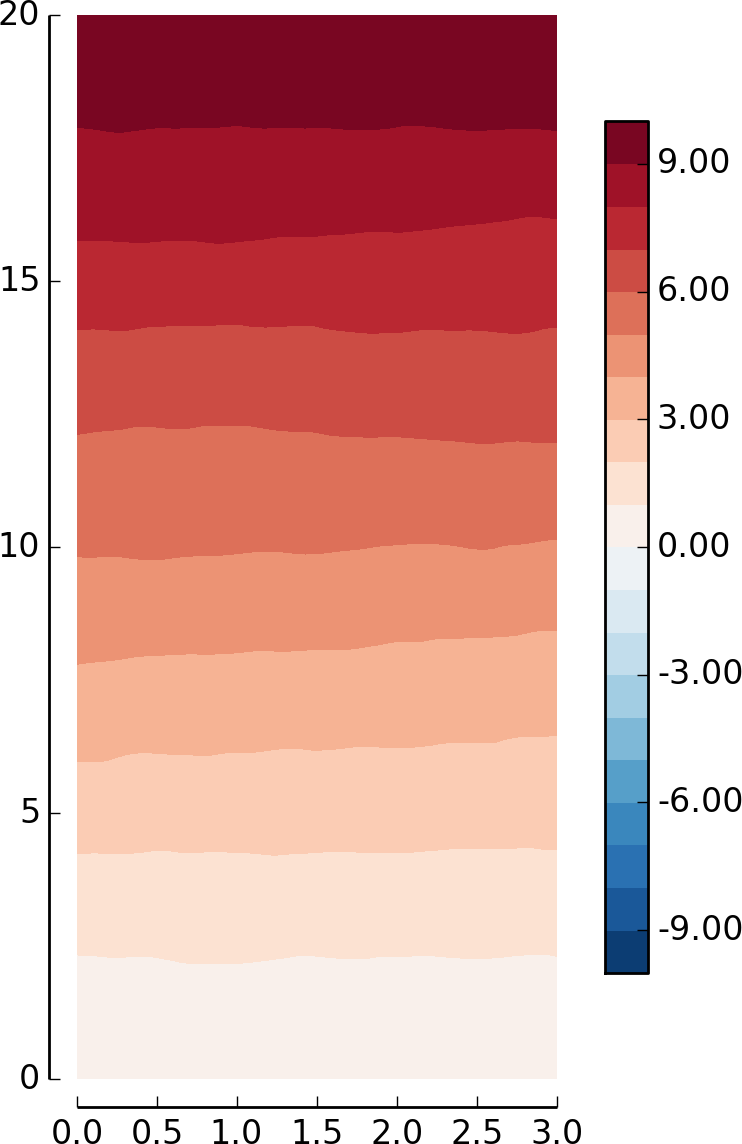}}\hfill%
	\subfigure[$\|{\Bu}\|$]{ 
	\includegraphics[height=0.45\textwidth]{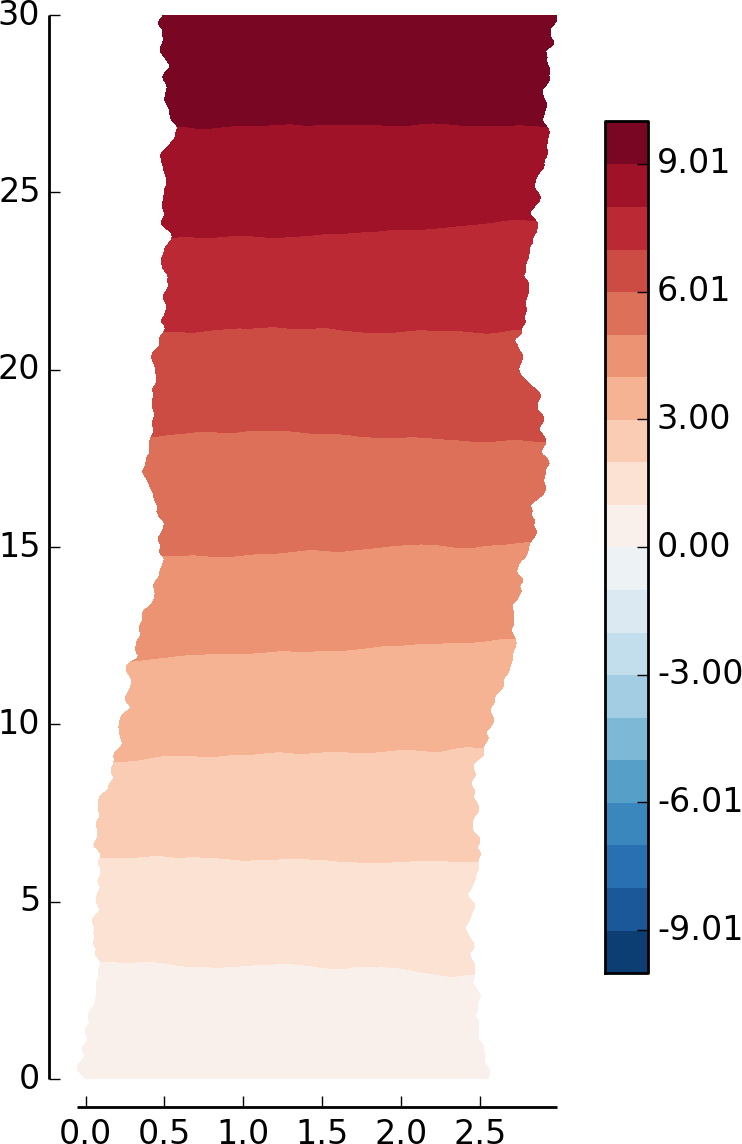}}
	\caption{	\label{fig:disp}
	Local field quantities obtained from a FEM simulation at maximum tensile 
	strain $\varepsilon_{yy}=0.5$: 
	(a) horizontal  displacements, (b) vertical displacements, 	(c) norm of the 
	displacements plotted in deformed geometry (note the different vertical length scale 
	as compared to plots (a) and (b)).}

\end{figure}

\begin{figure}
	\centering
	\subfigure[$\varepsilon_{yy}$]{%
	\includegraphics[height=0.45\textwidth]{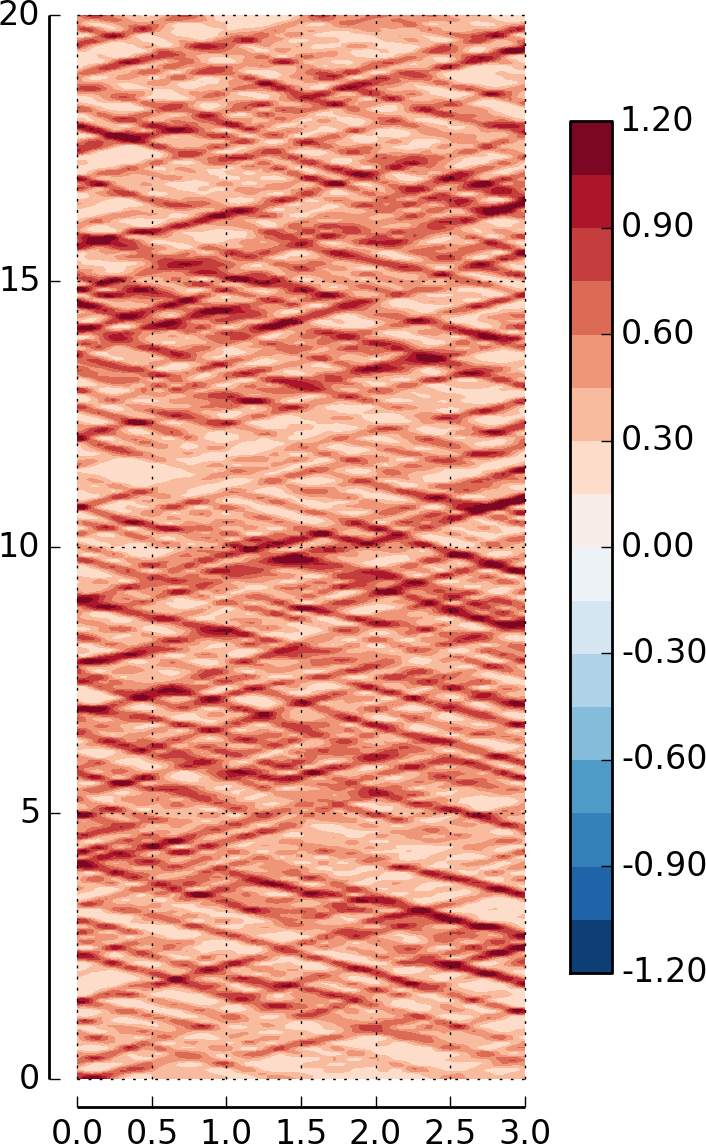}}%
	\qquad\quad
	\subfigure[$\varepsilon_{xy}$]{%
	\includegraphics[height=0.45\textwidth]{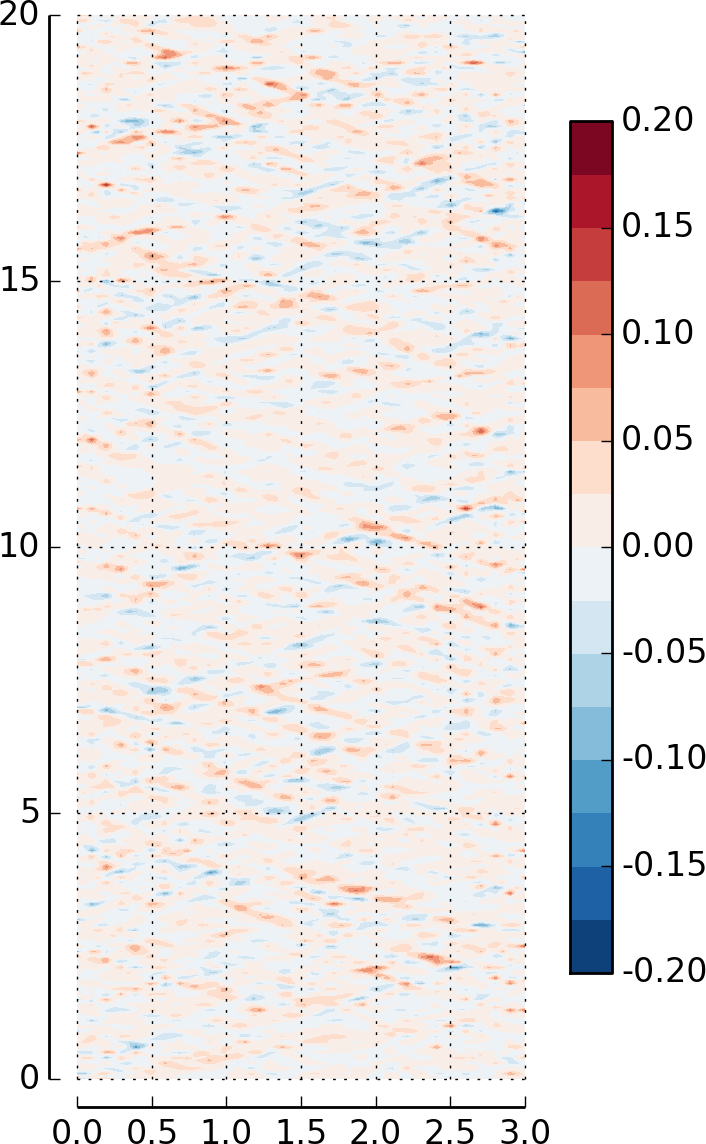}}%
	\caption{	\label{fig:strain}
	Examples of local strain distributions at maximum tensile 
	strain $\varepsilon_{yy}=0.5$: (a) axial strain $\varepsilon_{xx} \approx - \varepsilon_{yy}$ and 
	(b) shear strain $\varepsilon_{xy}$. Shear bands occur under an angle of $\pi/4$ w.r.t. the horizontal axis. 
	Note that the geometry is not true to scale.}
\end{figure}

For solution of the elasto-plastic boundary value problem we use the finite element method (FEM). The geometry is approximated by dividing the whole domain into $30$ by $200$ quadratic elements in $x$ and $y$ direction, respectively, so that the elements correspond to those used for defining the yield stress distribution. For approximation of the displacements we use quadratic shape functions and assume small strains and small rotations%
\footnote{This assumption is somewhat pushed to its limits in the following examples. It is, however, an acceptable approximation, since all important features such as shear bands and surface roughening already occur at early deformation stages and are merely amplified by increasing the total strain.}. %
Additionally, the prescribed vertical displacement $u_y$ is divided into small load increments.
As a result from the simulation we obtain the displacement field ${\Bu}=(u_x,u_y)$ as shown in \figref{fig:disp}. From the displacements one can derive the strain field as the symmetrical part of the displacement gradient, $\Bve=\text{sym}\left(\nabla {\Bu}\right)$, \figref{fig:strain}. 
The stress field can be incrementally obtained per load step from the strain field and is given by $\Delta{{\Bsigma}}=\IC^\text{ep}:\Delta{\Bve}$. Therein, $\IC^\text{ep}$ is the so-called elasto-plastic tangent modulus which simplifies to $\IC^\text{e}$ in case of a purely elastic load step (i.e., locally we have $f(\sigma_\text{eq})< 0$). In the case of plastic loading, $f(\sigma_\text{eq})=0$, $\IC^\text{ep}$ is the elastic modulus reduced by a plastic contribution, which depends on the hardening modulus $\Theta$  and which may thus change for each load increment due to a changed equivalent plastic strain (e.g. \cite{Simo1997}).

\begin{figure}
	\centering
	\includegraphics[width=0.95\textwidth]{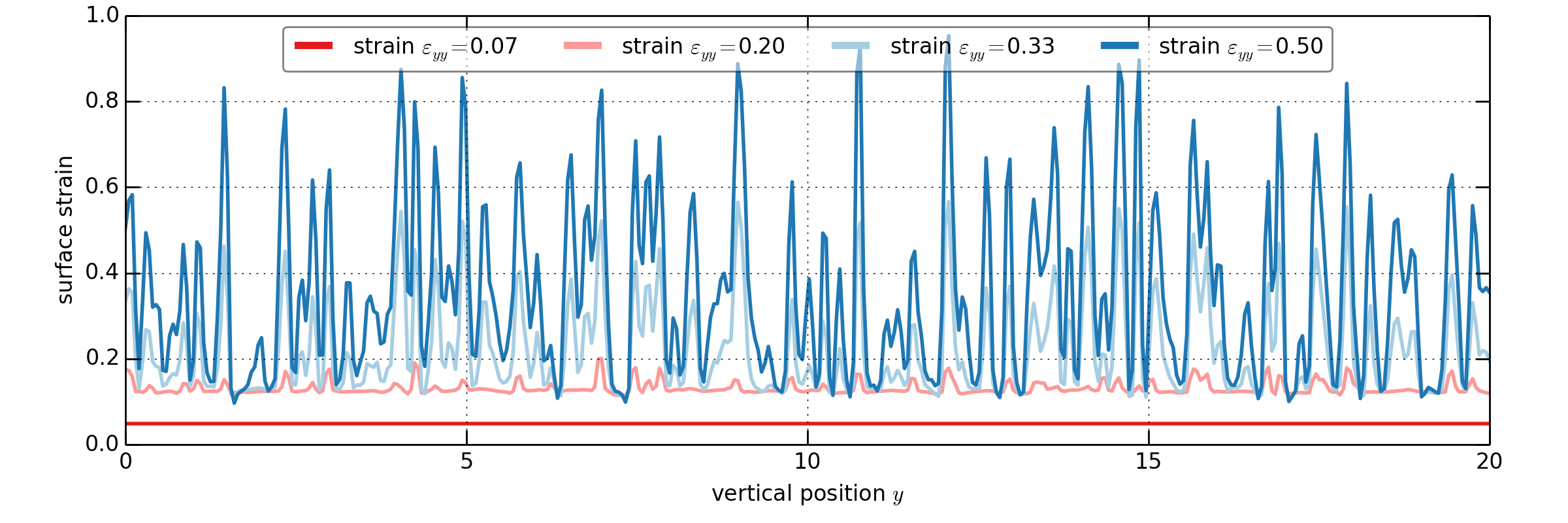}
	\caption{\label{fig:strain1d}
	Temporal evolution of the tensile strain at the left surface of the specimen shown in \figref{fig:strain}.}
\end{figure}

As shown in \figref{fig:strain}, axial strains are about one order of magnitude larger than shear strains, which indicates that multi-axiality of stress and strain does not have a strong influence on the deformation process. Accordingly, in the following we focus on the axial stress $\sigma_{yy}$ and axial strain $\varepsilon_{yy}$ for characterizing the deformation behavior. {An example of the evolution of the normal strain along the left surface is shown in \figref{fig:strain1d}. One observes that elements that have already undergone plastic deformation do  not revert to an elastic state, which is a particular feature of the $J_2$ plasticity model for the case of tensile loading. In fact, the principal features of the strain profile are established at an early stage after yield while further deformation simply makes the strain heterogeneities more pronounced without changing their spatial pattern.} 

\section{Stress strain curve and spatial fluctuations of stress and strain}
\label{sec:fluctuations}

\begin{figure}
	\centering
	\subfigure[average stress strain curve]{%
		\includegraphics[width=0.47\textwidth, viewport=0 0 338 250, clip]{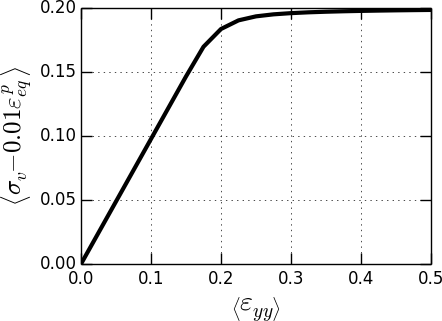}}%
		\;
	\subfigure[strain fluctuations]{%
		\includegraphics[width=0.47\textwidth, viewport=0 0 350 255, clip]{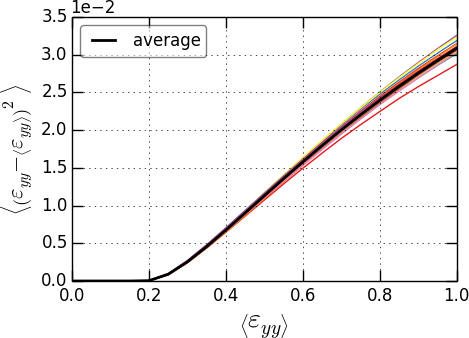}}\\%
	\subfigure[stress fluctuations]{%
		\includegraphics[width=0.47\textwidth, viewport=0 0 350 255, clip]{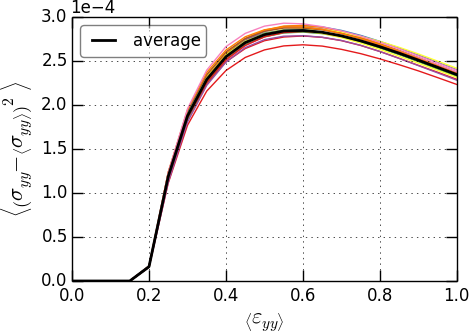}}%
		\;
	\subfigure[cross correlation]{%
		\includegraphics[width=0.47\textwidth, viewport=0 0 350 255, clip]{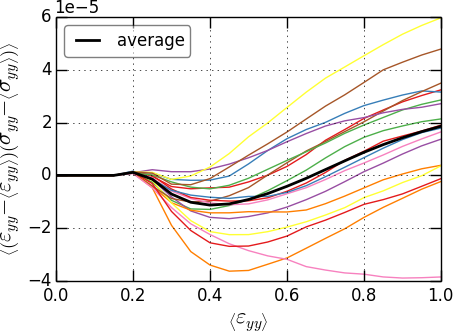}}\\
			
	\caption{\label{fig:fluctuations}
		(a) Stress strain curve, (b,c) local fluctuation amplitudes of local strain and stress, (d) cross correlation of local stress 
		and local strain; all variables are plotted as functions of the average axial strain. The solid lines denote averages over all
		simulations.}
\end{figure}

The global stress strain curves exhibit only very minor variations from specimen to specimen; \figref{fig:fluctuations} (a) shows therefore only the average stress strain response. Temporal fluctuations of the deformation rate (slip avalanches), which are a conspicuous feature of models where the flow stress fluctuates as a function of strain \cite{Talmali12,Talamali11,Zaiser05,Zaiser07}, are absent in the present model, and the main effect of the disorder on the stress strain curve is a rounded transition from the elastic to the plastic regime. Spatial fluctuations of stress and strain, on the other hand, arise naturally due to the random distribution of local flow stresses and form a conspicuous feature of the plastic flow regime which is characterized by localization of deformation into narrow shear bands (\figref{fig:strain}). We determine fluctuation magnitudes of the (axial) strain $\varepsilon_{yy}$ and (axial) stress $\sigma_{yy}$ as the respective differences between the local values of strain and stress on the finite element scale and their system-scale averages:
\begin{eqnarray}
\left\langle\delta \varepsilon_{yy}^2\right\rangle = \left\langle\left(\varepsilon_{yy}-\left\langle\varepsilon_{y}\right\rangle\right)^2\right\rangle
\qquad\text{and}\qquad
\left\langle\delta \sigma_{yy}^2\right\rangle =
\left\langle\left(\sigma_{yy}-\left\langle\sigma_{yy}\right\rangle\right)^2\right\rangle,
\end{eqnarray}
where $\langle\bullet\rangle$ denotes the average over the two-dimensional area. Fig.~\ref{fig:fluctuations} (b) and (c) show these quantities as functions of the average axial strain for individual samples. Fig.~\ref{fig:fluctuations} (d) additionally shows the cross correlation between stress and strain values, which is computed as
\begin{eqnarray}
\left\langle\delta \varepsilon_{yy}\delta \sigma_{yy}\right\rangle =
\left\langle \left(\varepsilon_{yy}-\left\langle\varepsilon_{yy}\right\rangle\right) \left(\sigma_{yy}-\left\langle\sigma_{yy}\right\rangle\right) \right\rangle.
\end{eqnarray}

Not surprisingly, the strain fluctuations increase with increasing deformation: after their first occurrence, strain localization patterns persist and become gradually more pronounced with increasing total strain. The initial increase can be described by a power law, $\langle \delta \varepsilon_{yy}^2 \rangle \propto (\langle \varepsilon_{yy} \rangle - \varepsilon_c)^{1.02}$ where $\varepsilon_c = 0.205$. Similarly, the stress fluctuations initially (i.e. up to $\ve_{yy}=0.45$) increase according to a power law, though with a much smaller exponent, $\langle \delta \sigma_{yy}^2 \rangle \propto (\langle \varepsilon_{yy} \rangle - \varepsilon_c)^{0.42}$.
The data are insufficient to decide whether, at larger degrees of deformation, these fluctuations saturate or continue to increase. Finally, the cross correlation between stress and strain exhibits large sample-to-sample variations. While initially, there is a clear and increasing negative correlation between stress and strain (locations with increased strain experience reduced stress and vice versa), which can again be described by a power law, with an exponent that is intermediate between those for the stress and strain autocorrelation, in later deformation stages stress and strain fluctuations become more and more decorrelated or may even exhibit positive correlations. We note that the power laws that may be used to describe the growth of fluctuations do not represent universal behavior, since the fitted exponents depend on the degree of disorder (a larger degree of disorder leads to a more gradual transition between the elastic and plastic regime). This is different for the statistical signatures of the strain localization patterns which we study in the following.

\section{Analysis of surface roughness and strain patterns}
\label{sec:surface_roughness}

\subsection{Root mean square of surface displacements}
\label{sec:rms}
To analyze the scale-dependent surface roughness we determine the root mean square (RMS) deviation of the surface displacements from their mean values for averaging windows of different length $l$. We move the averaging window from the bottom to the top of the sample and evaluate the average RMS for all possible positions of the averaging window. In the following graphs, we consider a set of 20 statistically equivalent simulations and analyze left and right surfaces separately, which gives a total number of $n = 40$ surfaces for ensemble averages.    

\begin{figure}[htb]
	\centering
  	\subfigure[using the narrow yield stress distribution $\sigma_\text{y0}=0.18\ldots 0.22$ ]{%
  	\includegraphics[width=0.47\textwidth, viewport=0 -10 350 500, clip]{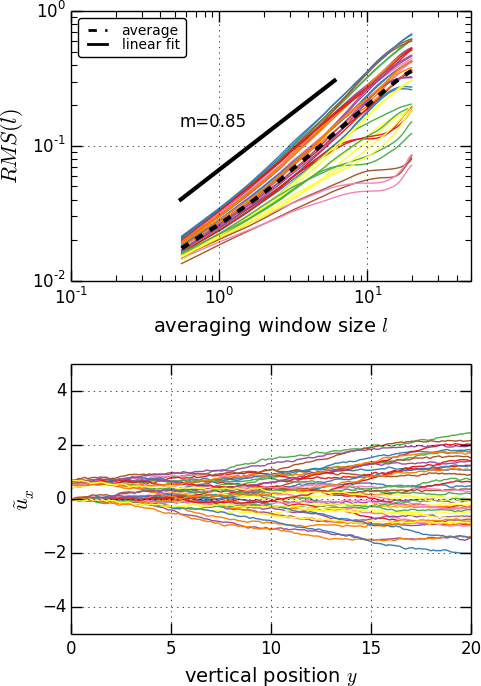}}
  	\quad
  	\subfigure[using the broader  yield stress distribution $\sigma_\text{y0}=0.05\ldots 0.35$]{%
  	\includegraphics[width=0.47\textwidth, viewport=0 -10 350 500, clip]{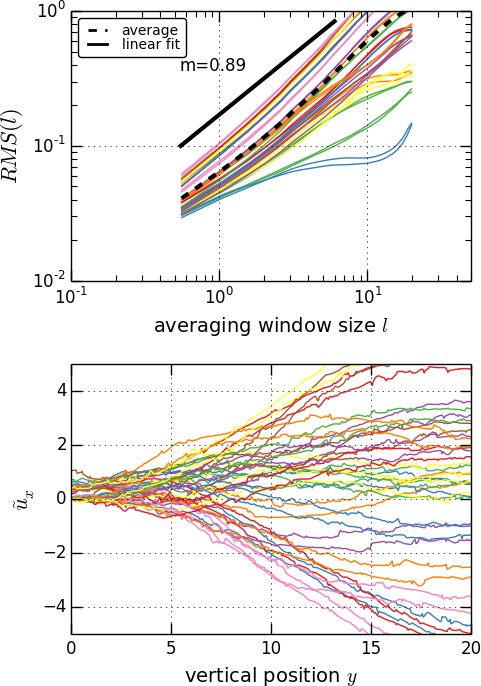}}
  	
	\caption{\label{fig:RMS} Horizontal surface displacements (bottom) and average RMS as a function of the averaging bin size $l$ (top) for yield stress distribution (A) and (B), at total strain $\varepsilon = 0.5$. The black dotted line shows the average of RMS$(l)$ over all simulations.}
\end{figure}

We denote by $\tilde u_x$ the horizontal surface displacement and by $\overline{u}_x(l,y)$ its average over an averaging bin of size $l$ located with its lower end at vertical position $y$:
\begin{equation}
	\overline{u}_x(l,y) := \langle \tilde u_x(y')\rangle_{y'}, \quad\text{where}\; y'\in[y\ldots y+l],
\end{equation}
where $\langle\bullet\rangle$ denotes the standard spatial average over different positions along the surface. The RMS is then given by
\begin{equation}
	\text{RMS}(l)= \left\langle \left\langle  \left(\tilde{u}_x(y')-\overline{u}_x(l,y)\right)^2  \right\rangle^{0.5}_{y'}  \right\rangle_{y}.
\end{equation}
where the subscript ${y}$ denotes the average over all $y$ with $0 \le y \le l_y - l$ on the given specimen surface. \figref{fig:RMS} shows the results for 20 simulations with two sets of surface data each. If we compare the surface displacement profiles for yield stress distributions with $\Delta\sigma=0.02$, (set A) with those obtained for a yield stress distributions with $\Delta\sigma=0.35$ (set B), we observe that the absolute value of the surface roughness is approximately twice as large for set (B) as compared to set (A) to (B). The scale dependence of RMS, however, is statistically equivalent in both cases as the ensemble-averaged RMS exhibits the same $l$ dependence: $\overline{\text{RMS}}(l) \propto l^H$ with an ensemble-averaged Hurst exponent of $0.85 \pm 0.05$ (set (A)) where the error of the ensemble averaged Hurst exponent equals the standard deviation of $H$ values determined for the single profiles, divided by $n^{1/2}$.

\subsection{Multi-scaling}
\label{sec:multiscaling}
In the following we analyze the surface roughening behavior in terms of the structure functions $S_{\alpha}(l)$ defined as
\begin{equation} \label{eq:multi_scaling}
  S_{\alpha}(l)=\Bigl\langle \left\vert \tilde u_x(y) - \tilde u_x(y+l)\right\vert^\alpha \Bigr\rangle^{1/\alpha}_{y}\;,
\end{equation}
where the subscript $y$ again denotes the average over all $y$ with $0 \le y \le l_y - l$. $\alpha$ is a positive parameter which we vary in the range $0.1 \le \alpha \le 30$. 
%
\begin{figure}[htb]
	\centering
  	\subfigure[$\alpha=1$]{
  	\includegraphics[width=0.48\textwidth, viewport=0 0 340 285, clip]%
  	{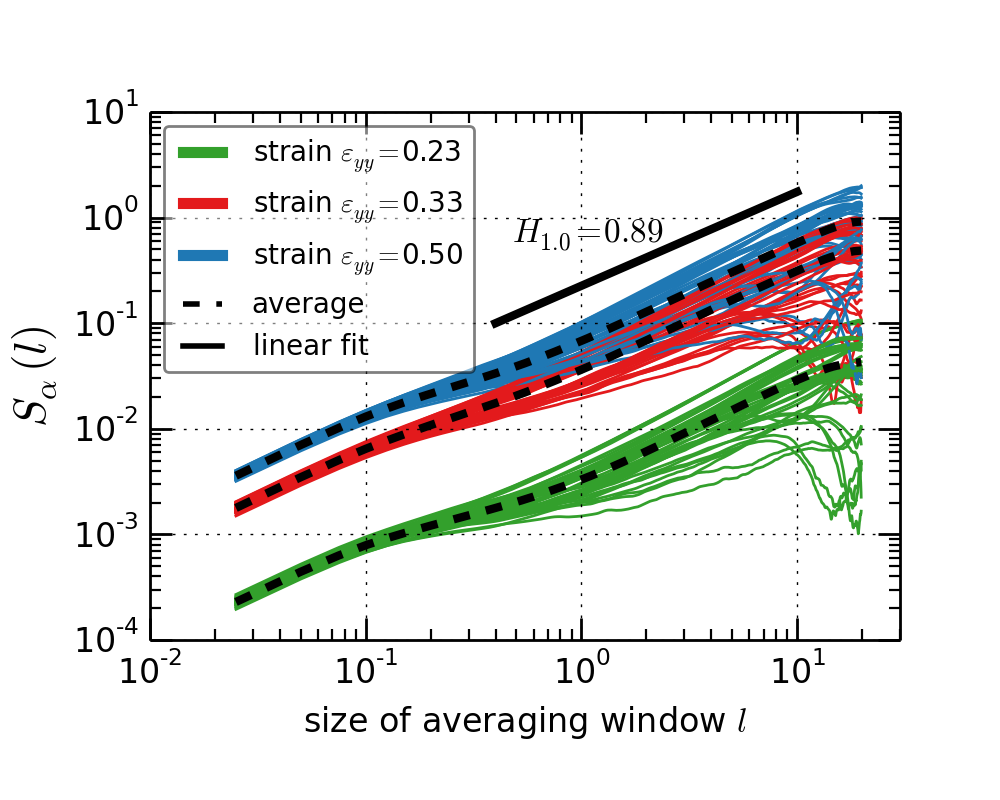}}\;
	\subfigure[fitted inclination as function of $\alpha$]{
	\includegraphics[width=0.48\textwidth, viewport=0 0 340 285, clip]%
	{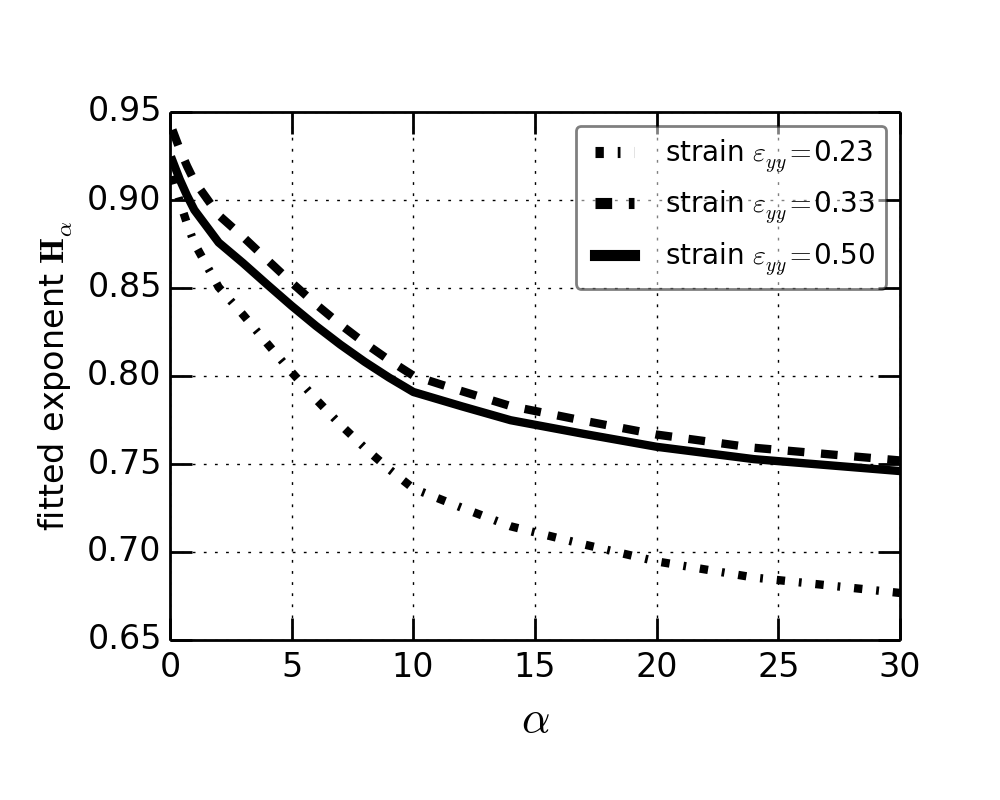}}
	\caption{\label{fig:Hl}%
Surface displacement structure functions; left: Structure function $S_1(l)$ for all simulated surfaces, right: exponent $H_{\alpha}$ as function of $\alpha$; both graphs show results for three different strain levels ($23\%$ total strain - shortly after yield, $33\%$, and in the final configuration at $50 \%$ total strain).}
\end{figure}
%
\figref{fig:Hl} (left) shows the shape of the structure functions $S_1(l)$ obtained from individual simulated surfaces at different stages of deformation, as well as the ensemble averaged structure functions (dashed lines). The curves can be divided into three different regions: (i) in the very left region ($l<10^{-1}$) we obtain a power law exponent close to $1$. In this region we are effectively below the resolution of the FEM discretization and do not measure the roughness of the surface but the 'roughness' of the FEM interpolation functions, yielding the value $H_{\alpha}=1$ expected for a smooth inclined surface. (ii) The second region ranges approximately from $l=10^{-1}\ldots 10^{+1}$ and can approximately be described by a power law, $S_{\alpha}(l) \propto l^{H_{\alpha}}$. This is indicated by the black, straight line which represents a linear fit over the interval $0.4 \le l \le 10$.  Each $H_{\alpha}$ point of \figref{fig:Hl}b represents the slope of such a fit, performed for the corresponding value $\alpha$ and at the respective total strain level. We find an absolute value of $H_1 = 0.89 \pm 0.05$ which is in good agreement with the Hurst exponent as determined from RMS values (the error of $H_1$ has again been determined from the standard deviation of fits to the structure functions of individual profiles). However, the multi-scaling analysis reveals that the exponents $H_{\alpha}$ decrease gradually w.r.t. $\alpha$, indicating that the surfaces are not self-affine but exhibit multiscaling properties. While the absolute values of the structure functions $S_{\alpha}$ increase continually with increasing strain, the surface morphology as characterized by the exponents $H_{\alpha}$ remains, outwith a narrow strain interval in the immediate vicinity of yield, unchanged throughout the plastic regime.

\subsection{Spatial correlation functions of local strain}
\label{sec:strain_corr}

To analyze shear band formation, i.e. strain localization occurring in preferred directions, we consider the spatial correlation function of the 
axial strain $\varepsilon=(-\varepsilon_{xx}+\varepsilon_{yy})/2$.  We evaluate the strain-strain correlation function as
\begin{equation}
	C_\varphi(l) := \Bigl\langle 
		\left(\varepsilon({\Br})         - \langle{\varepsilon}\rangle \right) 
		\left(\varepsilon({\Br}+\Bl_{\varphi}) - \langle{\varepsilon}\rangle \right) 
		 \Bigr\rangle_{\Br}.
\end{equation}
Here, the vector between two points, $\Br$ and $\Br+\Bl_{\varphi}$, is given in terms of its length $l=|\Bl|$ and angle $\varphi$ with the horizontal axis as $\Bl_{\varphi}:=[l\cos\varphi, l\sin\varphi]$. 

\begin{figure}[htb]
	\centering
  	\subfigure[$C_\varphi(l)$ plotted on linear axis]{
  	\includegraphics[height=0.37\textwidth, viewport=14 0 370 270, clip]%
  	{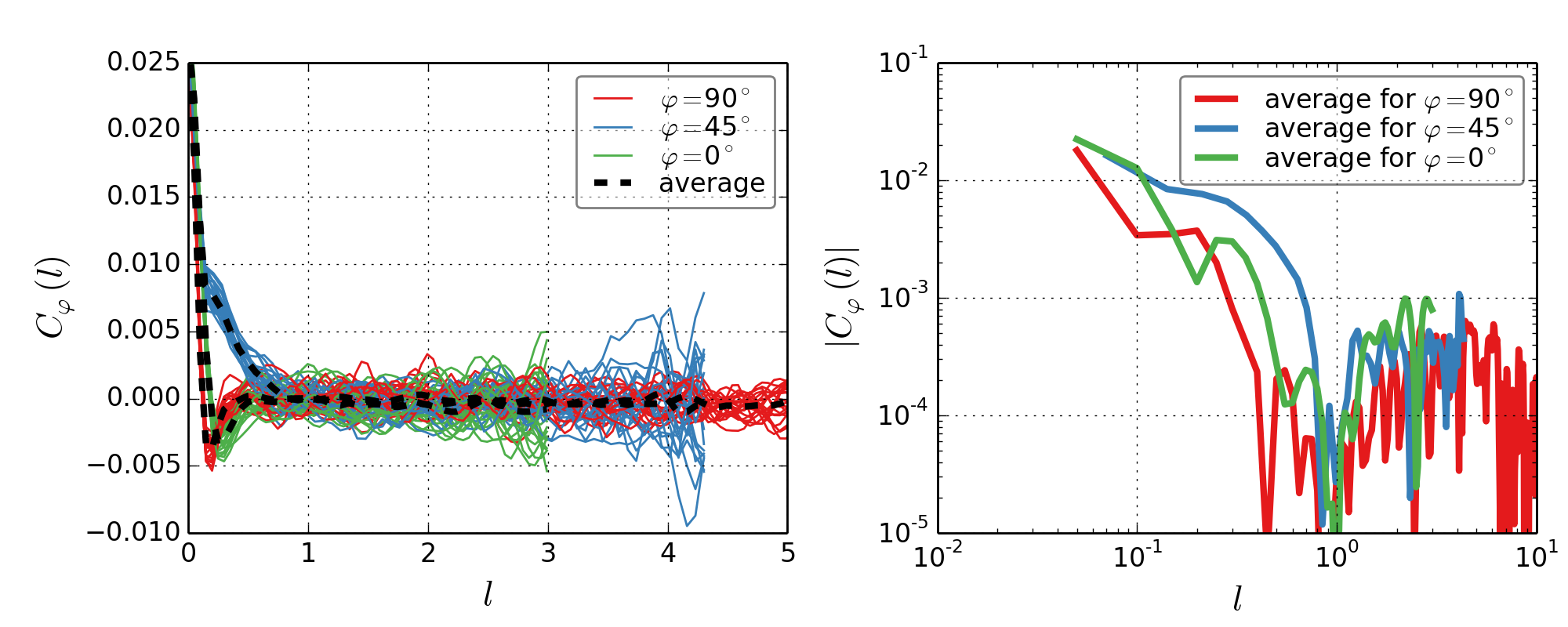}}
	\subfigure[absolute value of $C_\varphi(l)$ plotted on double logarithmic axis]{
	\includegraphics[height=0.37\textwidth, viewport=376 0 720 270, clip]%
	{figures/strain_corr1d.png}}\\
	\subfigure[two-dimensional strain correlation function]{
	\includegraphics[width=0.5\textwidth]%
	{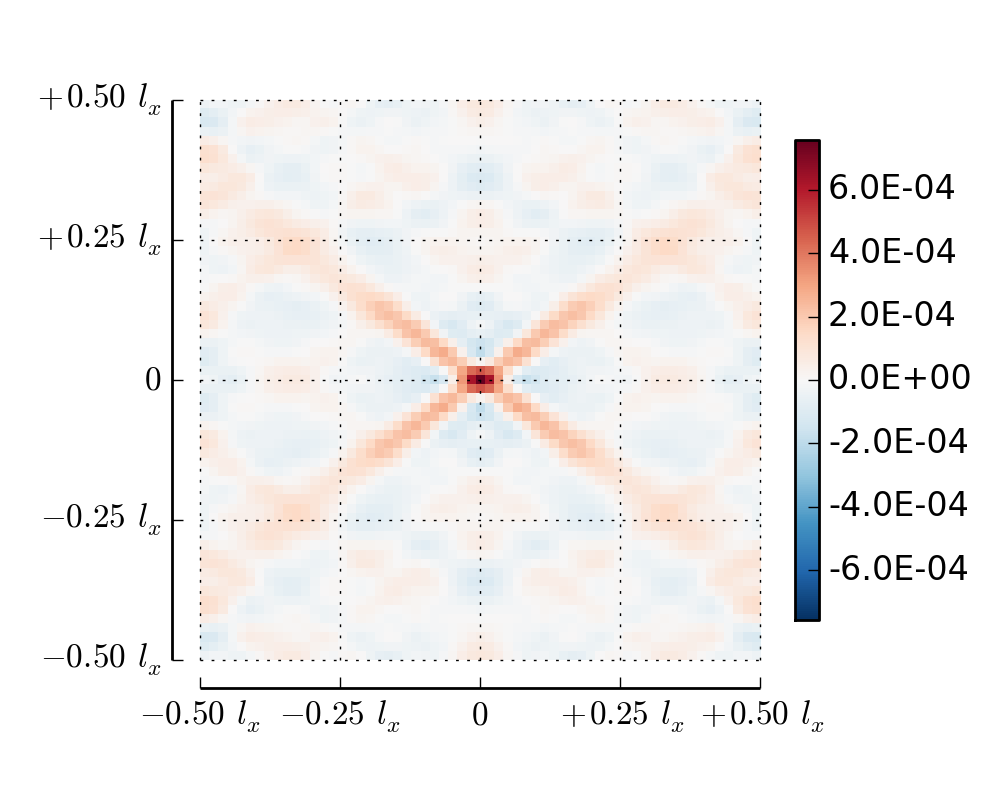}}
	\caption{\label{fig:strain_corr}%
		(a) and (b): strain correlation function as function of distance for different directions. The blue lines correspond the the diagonal direction of the shear bands. (c) shows the spatial correlation function.}
\end{figure}

A plot of the strain correlation function is shown in \figref{fig:strain_corr} (a) and (b). The directions $\varphi=0$ and $\varphi=\pi/2$ show positive correlations only on the scale of adjacent finite elements, followed by anticorrelations up to a range of about 5 elements and no statistically discernable correlation on larger scales. For the direction $\varphi=\pi/4$, which is the direction of the shear band formation, on the other hand we observe strong, long-ranged correlations extending from about 4 element lengths up to the specimen scale (blue lines). The double logarithmic plot indicates a tentative power law behavior $C_{\pi/4}(l) \propto l^{-\beta}$ with $\beta \approx 0.65$, even though the range and statistical quality of the data is not sufficient to unambiguously identify power law behavior. Fig.~\ref{fig:strain_corr} (c) shows the two-dimensional structure of the correlation function, where the $\pi/4$-direction can clearly be identified as the direction of shear bands. 
%

\section{Summary and Conclusions}

Our statistical analysis of a minimal model of amorphous plasticity indicates that formation of localized shear bands is a generic feature of disordered materials under plane strain or plane stress loading conditions. For the model under consideration, these shear bands are persistent features which become more pronounced with increasing total deformation, but do not diffuse in space. In this respect, the model is similar to the anti-plane shear model of Roux and Hansen \cite{Roux92} where deformation localizes into a single, system spanning band. This similarity can be directly related to the fact that both models assume the local yield stresses, while fluctuating in space, to be constant with strain. However, the different deformation geometry (plane vs. anti-plane shear) gives rise to a very different slip band morphology. While in the case of anti-plane shear, Roux and Hansen \cite{Roux92} find a single, system spanning band with self-affine morphology, we observe multiple slip bands which simply follow the 45$^o$ direction of maximum shear stress . At the same time, we observe non-trivial surface roughness with a Hurst exponent around 0.8 which indicates the presence of subtle, long-range correlations between these bands. Long range correlations are also evidenced by a slow decay of the 2D strain fluctuation correlation function in the slip band direction, though a quantitative analysis of the spatial decay law was not possible due to the limited statistics available. 

The long-range correlations of the incipient surface roughness are already present at a very early stage of deformation. The roughening process is thus very different from roughening phenomena associated with interface motion or growth processes, where roughness emerges first on small scales and then gradually extends to larger and larger ones. In the present case, non-trivial roughness arises from the interplay between local disorder and long-range stress re-distribution. Even on large scales, this roughness is present from the onset of plastic deformation: the pattern of surface height fluctuations, while growing in amplitude, remains largely stationary in space. The persistence of the localization and roughness patterns observed in our model is consistent with the behavior observed experimentally in polycrystals \cite{Wouters05,Wouters06} -- in fact, the present model may provide a good physical approximation of poly- and nanocrystalline materiasls where plasticity can be approximated as isotropic, while persistent, strain independent fluctuations in yield stress arise from the initial scatter of grain orientations. 

The Hurst exponents determined from our simulations are in agreement with those determined from other models \cite{Zaiser05} and from experiment \cite{Zaiser04,Wouters05,Schwerdtfeger10}. It is particularly interesting to compare the present findings with those derived from the simulations by Zaiser and Moretti \cite{Zaiser05} and with the observations from the experimental work of Schwerdtfeger et. al. \cite{Schwerdtfeger10}. These studies refer to single crystals where disorder arises from dislocation interactions and therefore evolves with strain. This leads to non-stationary, diffusing strain localization patterns, i.e., the evolution is fundamentally different from the present model. Nevertheless, the Hurst exponents determined are similar to the present case, which provides some indication that the interplay between disorder and long-range elasticity may lead to generic, universal features of the resulting strain localization patterns and surface morphologies.

\section*{Acknowledgments}
MZ gratefully acknowledges financial support by EPSRC under grant no. EP/J003387/01.\\


\section*{References}
\begin{thebibliography}{}
%
%

\bibitem{Steif82} P.S. Steif, F. Spaepen and J.W. Hutchinson, Acta Metall. 30, 447 (1982).
\bibitem{Poliakov94} A.N.B. Poliakov, H.J. Herrmann, Y.Y. Podladchikov and S. Roux, Fractals 2, 567 (1994).  
\bibitem{Sun11} B. A. Sun and W. H. Wang, Appl. Phys. Lett. 98, 201902 (2011).
\bibitem{Roux92} S. Roux and A. Hansen, J. Phys. II (France) 2, 1007 (1992). 
\bibitem{Talmali12} M. Talamali, V. Petäja, D. Vandembrouq and S. Roux, Comptes-Rendus Mécanique 340, 
     275 (2012). 
\bibitem{Talamali11} M. Talamali, V. Petäja D. Vandembroucq and S. Roux, Phys. Rev. E 84, 016115 (2011).
\bibitem{Zaiser05}  M. Zaiser and P. Moretti J. Stat. Mech., P08004 (2005).
\bibitem{Zaiser07}  M. Zaiser and P. Nikitas, J. Stat. Mech., P04013 (2007).
\bibitem{Zaiser04}  M. Zaiser, F.M. Grasset, V. Koutsos and E.C. Aifantis, Phys. Rev. Letters  93, 195507 (2004).
\bibitem{Wouters05} O. Wouters, W. Vellinga, R. Tijum and J. de Hosson, Acta Mater. 53, 4043 (2005). 
\bibitem{Wouters06} O. Wouters, W. Vellinga, R. Tijum and J. de Hosson, Acta Mater. 54, 2813 (2006).
\bibitem{Schwerdtfeger10}  J. Schwerdtfeger, E. Nadgorny, V. Koutsos, J.R. Blackford and M. Zaiser, Acta Mater. 58, 4859 (2010).
\bibitem{Kleiser86} T. Kleiser and M. Bocek, Z. Metallkd. 77, 582 (1986).
\bibitem{Burnley2013} P. C. Burnley, Nature Communications, 4:2117 (2013)
\bibitem{Simo1997} J.C. Simo and T.J.R. Hughes, Computational Inelasticity, Springer New York (1997)
\end{thebibliography}


\end{document}